\documentclass[12pt]{article}
\usepackage{graphicx}
\pdfoutput=1
\newcommand\pubnumber{DPF2015-384}
\newcommand\pubdate{\today}
\newcommand{\numu}{$\nu_\mu$~}
\newcommand{\numus}{$\nu_\mu$s~}
\newcommand{\numubar}{$\bar{\nu}_\mu$~}
\newcommand{\nue}{$\nu_e$~}
\newcommand{\nuebar}{$\bar{\nu}_e$~}
\def\MIT{Laboratory for Nuclear Science\\
The Massachusetts Institute of Technology, Cambridge, MA 02139}
\def\AA{Department of Physics\\
University of Michigan, Ann Arbor, MI 48109}
\def\NEVIS{Nevis Laboratories\\
Columbia University,  New York, NY 10027}
\def\support{\footnote{email: saxani@mit.edu}}
\def\Title#1{\begin{center} {\Large #1 } \end{center}}
\def\Author#1{\begin{center}{ \sc #1} \end{center}}
\def\Address#1{\begin{center}{ \it #1} \end{center}}

\newcommand\pubblock{\rightline{\begin{tabular}{l} \pubnumber\\
         \pubdate  \end{tabular}}}
\newenvironment{Abstract}{\begin{quotation}  }{\end{quotation}}
\newenvironment{Presented}{\begin{quotation} \begin{center} 
             PRESENTED AT\end{center}\bigskip 
      \begin{center}\begin{large}}{\end{large}\end{center} \end{quotation}}

\def\beq{\begin{equation}}
\def\eeq#1{\label{#1}\end{equation}}
\def\eeqn{\end{equation}}

\def\beqa{\begin{eqnarray}}
\def\eeqa#1{\label{#1}\end{eqnarray}}
\def\eeqan{\end{eqnarray}}

\let\bar=\overbar

\def\Dslash{\not{\hbox{\kern-4pt $D$}}}
\def\dslash{\not{\hbox{\kern-2pt $\del$}}}

\def\msb{{\bar{\ssstyle M \kern -1pt S}}}

\begin{document}
\begin{titlepage}
\pubblock
\vfill
\Title{KPipe: a decisive test for muon neutrino disappearance}
\vfill
\Author{ S.N. Axani\support,  G. Collin, J.M. Conrad, T. Wongjirad}
\Address{\MIT}
\Author{M.H. Shaevitz}
\Address{\NEVIS}
\Author{and J. Spitz}
\Address{\AA}
\vfill
\begin{Abstract}
The short baseline neutrino oscillation experiment, KPipe, is designed to perform a sensitive search for muon neutrino disappearance in the current global fit allowed regions for sterile neutrinos. KPipe is to be located at the Material Life Science Experimental Facility at J-PARC: the world's most intense source of 236~MeV, monoenergetic muon neutrinos. By measuring the \numu charged current interaction rate along a 120~m long, 3~m diameter detector, KPipe can map out short baseline oscillations over an L/E of 0.14 to 0.64~m/MeV. Using a long, single detector to measure the \numu interaction rate as a function of distance largely eliminates the systematic uncertainties associated with cross sections and fluxes. In this paper, we show that KPipe can cover the current global best fit to 5$\sigma$ after 3 years of running.
\end{Abstract}
\vfill
\begin{Presented}
DPF 2015\\
The Meeting of the American Physical Society\\
Division of Particles and Fields\\
Ann Arbor, Michigan, August 4--8, 2015\\
\end{Presented}
\vfill
\end{titlepage}
\def\thefootnote{\fnsymbol{footnote}}
\setcounter{footnote}{0}
\section{Introduction}
Several short baseline (SBL) neutrino experiments have observed anomalies consistent with the signal expected by a new, non-weakly interacting (sterile) neutrino state with a $\Delta$m$^2\approx1$~eV$^2$. These anomalies include the observation of an excess number of \nuebar events in a \numubar beam from LSND~\cite{Aguilar:2001ty}, an excess in \nue and \nuebar events from a \numu and \numubar beam in MiniBooNE~\cite{AguilarArevalo:2008rc,AguilarArevalo:2007it}, and a deficit of \nuebar events from the global reactor~\cite{Zhang:2013ela} and intense radioactive source experiments~\cite{Giunti:2010zu}. However, due to the null-appearance measurement from the KARMEN experiment~\cite{karmen}, cosmological constraints \cite{cosmo}, and the lack of observation regarding moun neutrino disappearance~\cite{muon}, there is significant tension~\cite{Ignarra:2014yqa} within the sterile neutrino paradigm. There are mechanisms to evade the null-oscillation result and cosmological constraints \cite{white}, but if there exists a light sterile neutrino, there must be some amount of muon neutrino disappearance. In contrast, the null-measurement of muon neutrino disappearance would help constrain the interpretation of the oscillation signals seen in LSND and MiniBooNE.

In the short baseline approximation, m$_1\approx$~m$_2\approx$~m$_3\gg$~m$_4$, where m$_{1,2,3}$ represent the active neutrino masses and m$_4$ sterile neutrino mass, the probability for muon neutrino disappearance is:
\begin{equation}
P(\nu_{\mu}\rightarrow\nu_{\mu})\simeq
1-4(1-|U_{\mu 4}|^2)|U_{\mu 4}|^2\sin^2(1.27\Delta m^2_{41}L/E)~, \label{disappeq1}
\end{equation}
where $U_{\mu 4}$ establishes the matrix element between the muon neutrino state and the sterile neutrino state, L is the baseline, E is the neutrino energy, and $\Delta m^2_{41}$ is the mass square splitting between the active and sterile mass-eigenstates. 

The following describes how the SBL neutrino oscillation experiment, KPipe, can address the muon neutrino disappearance tension by performing a precision measurement of the rate of charged current (CC) interactions induced by 236~MeV, monoenergetic \numus as a function of distance. We show that based on a shape-only analysis of the oscillation wave through the detector, KPipe can extend the experimental exclusion limits an order of magnitude and measure the current global best fit 5$\sigma$ after 3 years of running.

\section{The Experiment}
\subsection{The Source}
The Japan Proton Accelerator Research Complex (J-PARC) Materials and Life Science Experimental Facility (MLF) neutron spallation source was built to provide intense neutron and muon beams for material probing by colliding a 3 GeV proton beam onto a mercury target station. The beam structure is designed to deliver protons in pairs of bunches, spaced 80 ns apart at repetition rate of 25~Hz. Recently~\cite{power}, the MLF has demonstrated that it can operate at the final design power of 1~MW, which, with 5000 hours/year \cite{jaea}, corresponds to 3.75$\times 10^{22}$~POT/year. The interaction between the high-intensity proton beam and a mercury target station produces an abundance of neutrons and other secondary particles, including kaons, pions, and muons. Of particular interest to KPipe are the K$^+$s, which preferentially decay-at-rest in the mercury target to a $\mu^+$ and a 236~MeV, monoenergetic \numu (K$^+\rightarrow  \mu^+\nu_\mu$ at a branching ratio of 63.55 $\pm$ 0.11\%~\cite{pdg}). This is the world's most intense source of KDAR neutrinos, which are important for multiple physics measurements~\cite{kdar1,kdar2}, including the oscillation search described here.

\subsection{The KPipe anti-neutrino detector and signal} \label{sec:detector}

\begin{figure}[t!]
\centering
   	 \begin{minipage}{.48\linewidth}
		\includegraphics[width=3.0in]{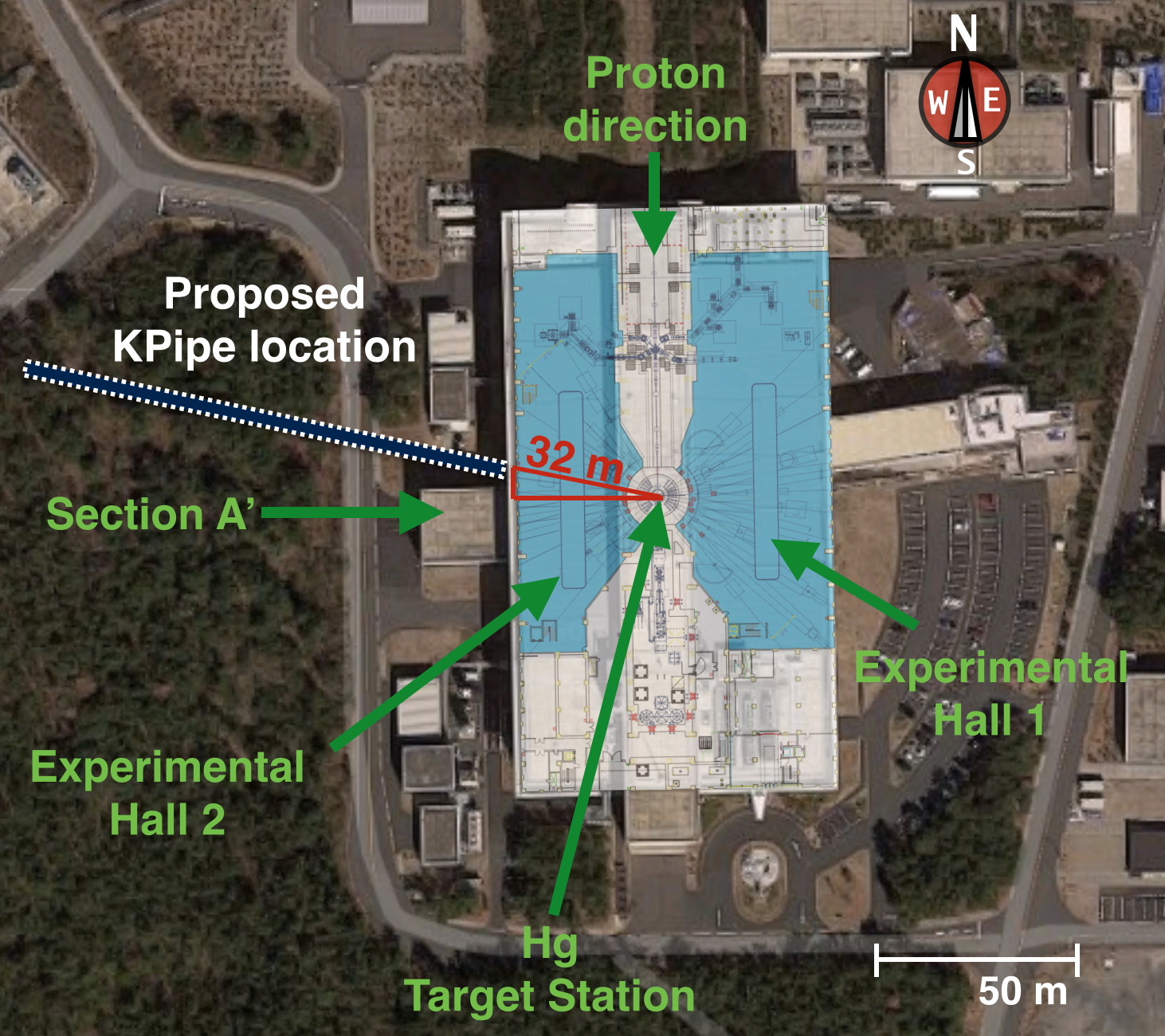}
    \end{minipage}
    \hspace{.01\linewidth}
    \begin{minipage}{.48\linewidth}
		\includegraphics[width=3.0in]{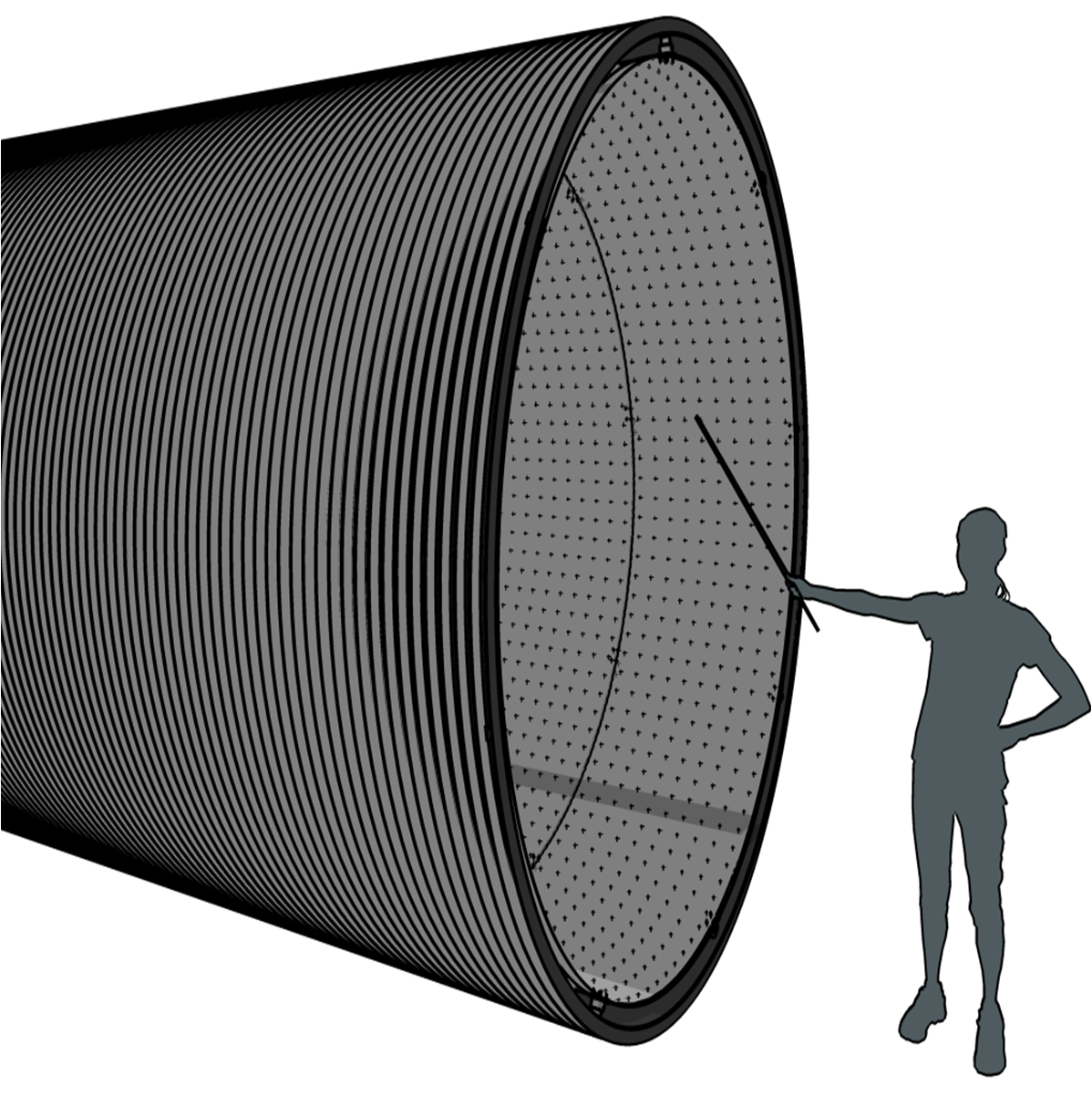}
    \end{minipage}
\caption{Left:  An aerial view of the MLF and the proposed KPipe location. Right: A rendering of the KPipe detector.}
\label{fig:mlf}
\end{figure}

KPipe can utilize the source of monoenergetic $\nu_\mu$s by measuring the \numu CC interaction on the carbon nuclei of a liquid scintillator (LS) to perform a sensitive SBL oscillation measurement. Figure \ref{fig:mlf} (left) shows an aerial view of the MLF and the proposed location for the KPipe detector. This particular location was chosen to maximize the sensitivity to SBL neutrino oscillations while being as non-obstructive to the MLF as possible. KPipe is a single, 120~m long, 3~m diameter detector, constructed out of high-density polyethylene (HDPE) and extends radially outwards at cos$\theta=-0.21$~(102$^\circ$) with respect to the proton direction. It's closest point lies 32~m from the target station and therefore is able to measure L/E from 0.14 to 0.64~m/MeV. It will be buried several meters below the ground or housed in a long concrete building to provide some overburden. Using a single detector eliminates the systematic uncertainties associated with interaction cross sections and production rates, and reconstructing an oscillation wave as a function of distance can provide conclusive evidence for neutrino oscillation.

The detector is divided into an outer volume (the cosmic ray (CR) veto) and an inner volume (the target volume), as shown in Figure \ref{fig:mlf} (right). The two volumes are optically separated by 1.45~m radius, highly reflective, cylindrical panels. On the interior of the panels, 6x6~mm silicon photomultipliers (SiPMs) are arranged in hoops and point radially inwards into the 793 m$^3$ (683 tonne) fiducial volume. Each hoop contains 100 equally spaced SiPMs and each hoop is spaced by 10~cm along the length of the detector. This provides a total photocathode coverage of 0.4\%. On the exterior of the panels, there are another 120 hoops, on which are mounted 100 equally spaced SiPMs, and face radially outward into the CR veto. 

The CC interaction between the \numu and the carbon nuclei in the LS (\numu $^{12}$C~$\rightarrow~\mu^-$X, where X represents a number of final state probabilities) can be identified by the scintillation signal from the prompt $\mu^-$ and final state proton, and the subsequent delayed signal from the Michel electron ($\mu^-\rightarrow\bar{\nu}_ee^-\nu_\mu$). 

\section{The simulations and signal selection}

\begin{figure}[!t]
\centering
   	 \begin{minipage}{.48\linewidth}
		\includegraphics[width=2.90in]{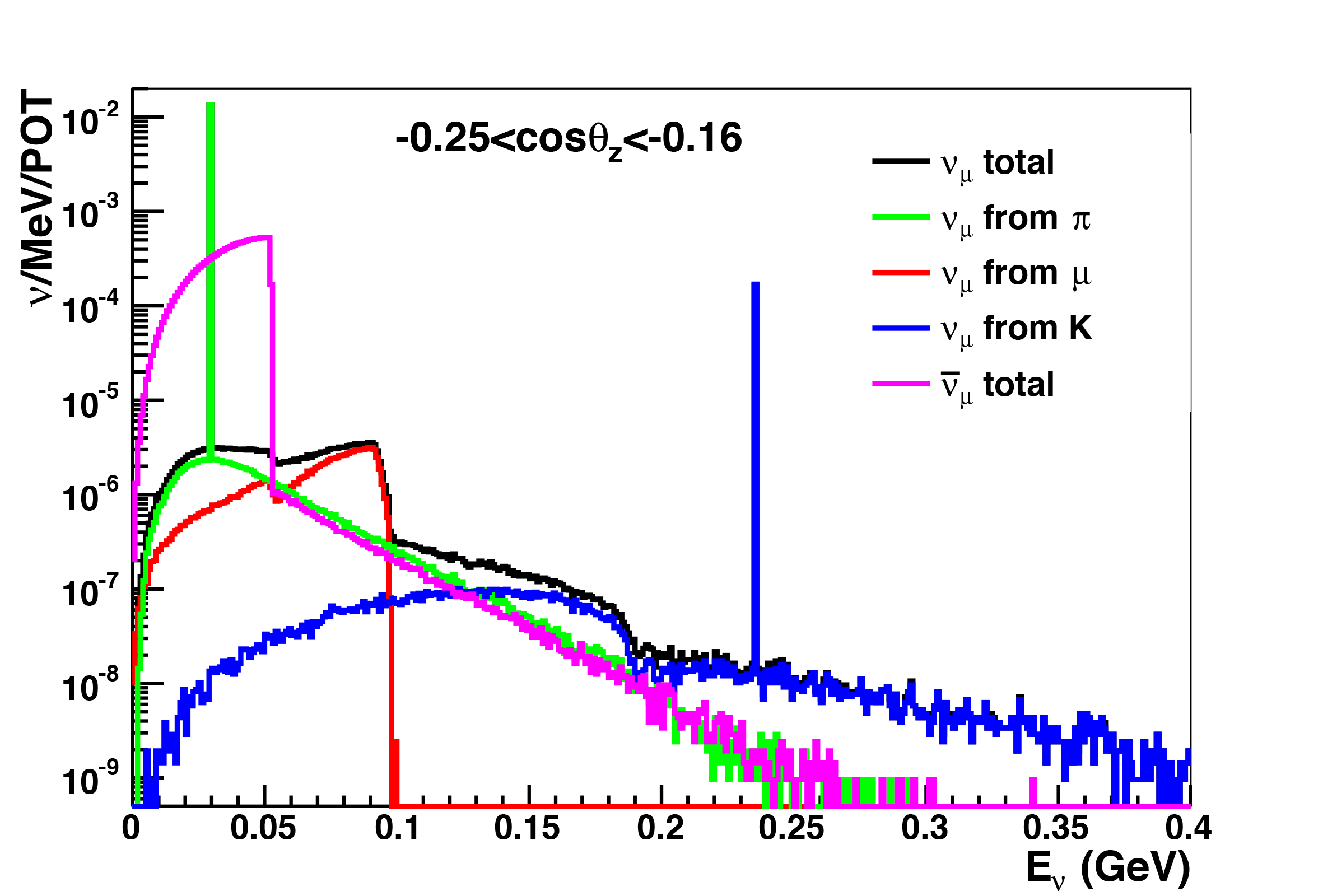}
    \end{minipage}
    \hspace{.01\linewidth}
    \begin{minipage}{.48\linewidth}
		\includegraphics[width=2.90in]{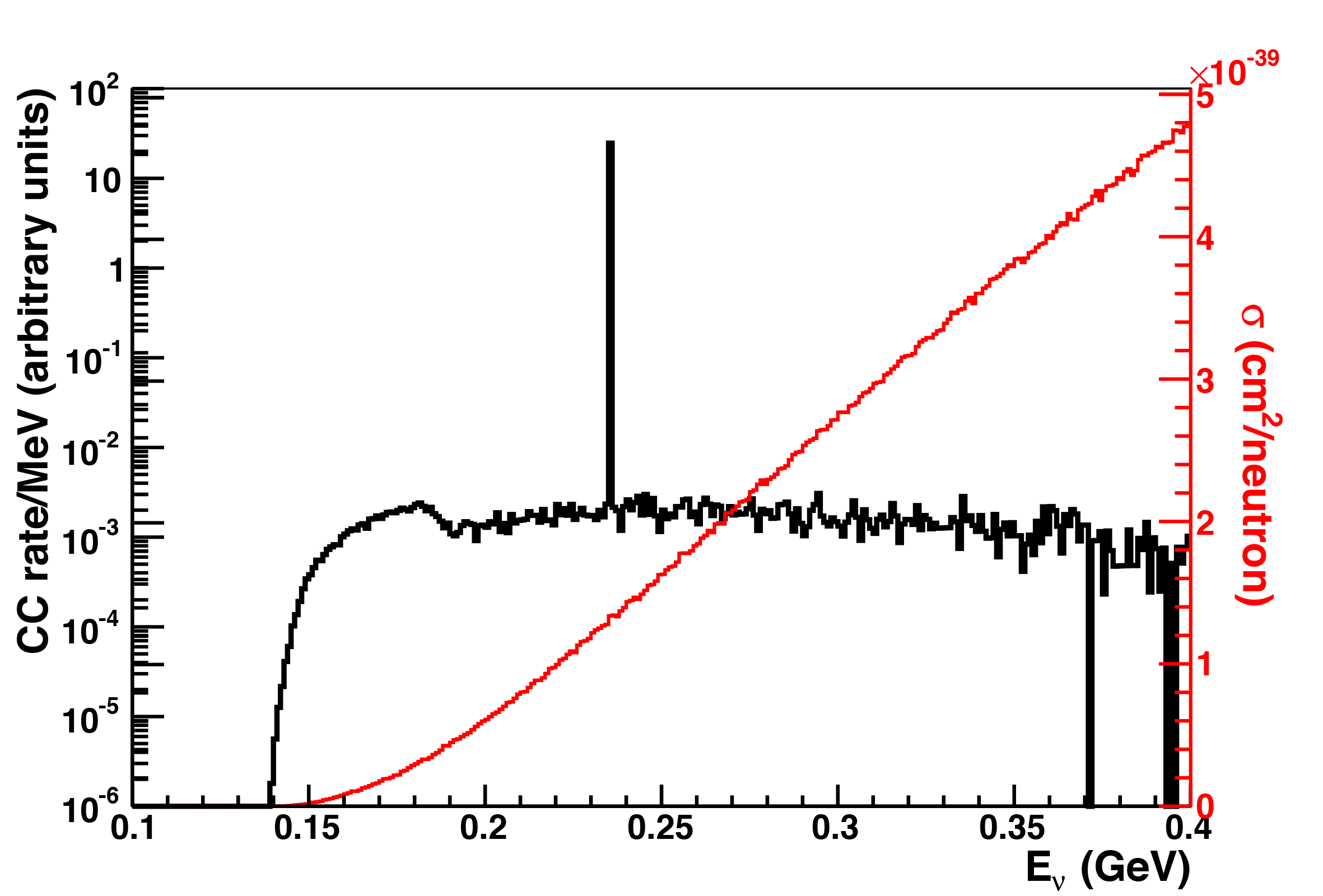}
    \end{minipage}
\caption{Left: The relevant predicted \numu flux, calculated by Geant4, as a function of energy through the solid angle $-0.25<\mathrm{cos}\theta<-0.16$. Right: The \numu CC interaction cross section on a neutron (red), and the corresponding interaction rate (black).}
\label{fig:rate}
\end{figure}

The neutrino flux from the incident 3 GeV protons on the mercury target was determined using a combination of Geant4~\cite{Geant4} and MARS15~\cite{mars}. Using a semi-realistic model of the mercury target in Geant4, approximately 92\% of the K$^+$s were found to DAR, 75\% of which were found to be located within 25~cm of the K$^+$ production point. MARS15 was used to calculate the relative muon neutrino flux and energy associated with particles emitted from the source. This is shown in Figure \ref{fig:rate} (left). In particular, the MARS15 model predicts the 236~MeV \numu production rate to be 0.0072~\numu/POT. 

The interaction of the neutrinos with the LS and the kinematics of the final state particles were modeled using the event generator NuWro~\cite{NuWro}. The \numu CC cross section and interaction rate per neutron in the LS is shown in Figure \ref{fig:rate} (right). The KDAR \numu interaction rate purity is found to be 98.5\%, with a cross section per neutron of 1.3$\times 10^{-39}$ cm$^2$. Therefore, we expect 1.02~$\times~10^5$ KDAR \numu CC events/year over the entire detector fiducial volume, 98.5\% of which come from a 236~MeV $\nu_\mu$.

To determine the detection efficiency and CR induced interaction rate in the detector and surrounding material, the Geant4 based simulation package RAT~\cite{rat} was used. We also used CRY~\cite{cry} to predict the relative amount of munos, pions, electron, photons, neutrons, and protons in the CR flux. The RAT model of the detector incorporates 130,200, 6x6 mm$^2$ SiPMs, each with a 1.6 MHz dark rate at a quantum efficiency of 30\%, highly reflective interior walls, and LS with a conservative light yield of 4500 photons/MeV. Based on the simulated signals produced by the prompt $\mu^-$ and the delayed Michel electron, we find that 87\% of the muon from the \numu CC interaction are contained within the fiducial volume; 99.5\% of the muons are identified above the SiPM noise; and 88.5\% of the subsequent Michel electrons can be identified (after energy, spatial coincidence, and timing cuts are applied). Based on the number of photoelectrons detected in the SiPMs, we find that we can reconstruct the vertex of the interaction point to within 0.8~m.

The primary background to the \numu CC interaction detection was found to originate from stopping cosmogenic muons. A high-energy cut removed 72\% of the total number of CR induced events, and a low-energy cut (for secondary spallation products and primary CRs) removed another 15\%. Finally, the 10~cm veto was found to remove 99.5\% of the remaining CR flux, leaving a remaining rate of 27 Hz along the total detector. The coincidence between a CR interaction in the veto and a \numu CC event was found to occur 2.6\% of the time. This, combined with the signal efficiency described in the previous paragraph, corresponds to a total signal efficiency of 75\%.

\section{Sensitivity studies}
\begin{figure}[t]
\centering
   	 \begin{minipage}{.48\linewidth}
		\includegraphics[width=2.8in]{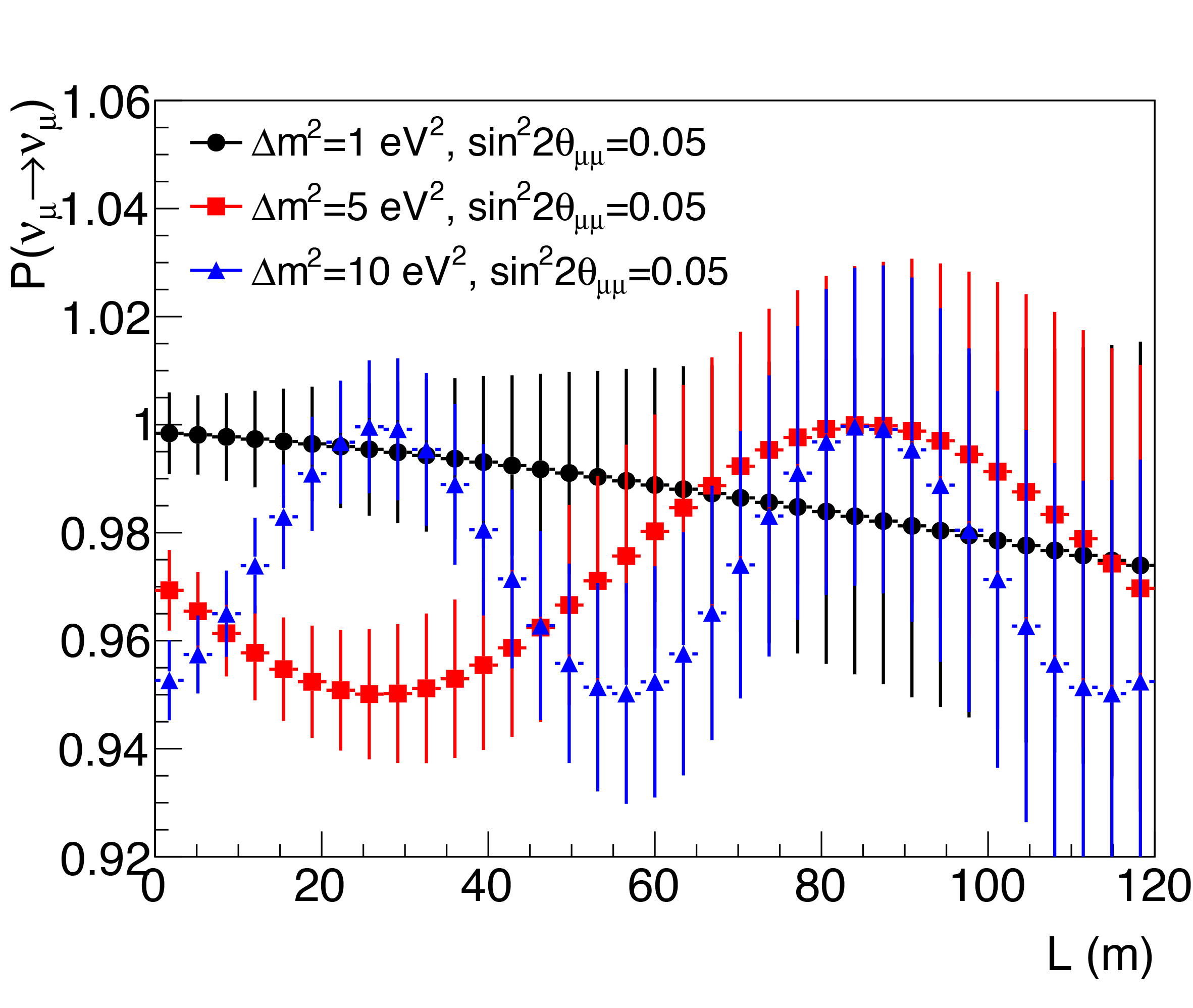}
    \end{minipage}
    \hspace{.01\linewidth}
    \begin{minipage}{.48\linewidth}
		\includegraphics[width=2.8in]{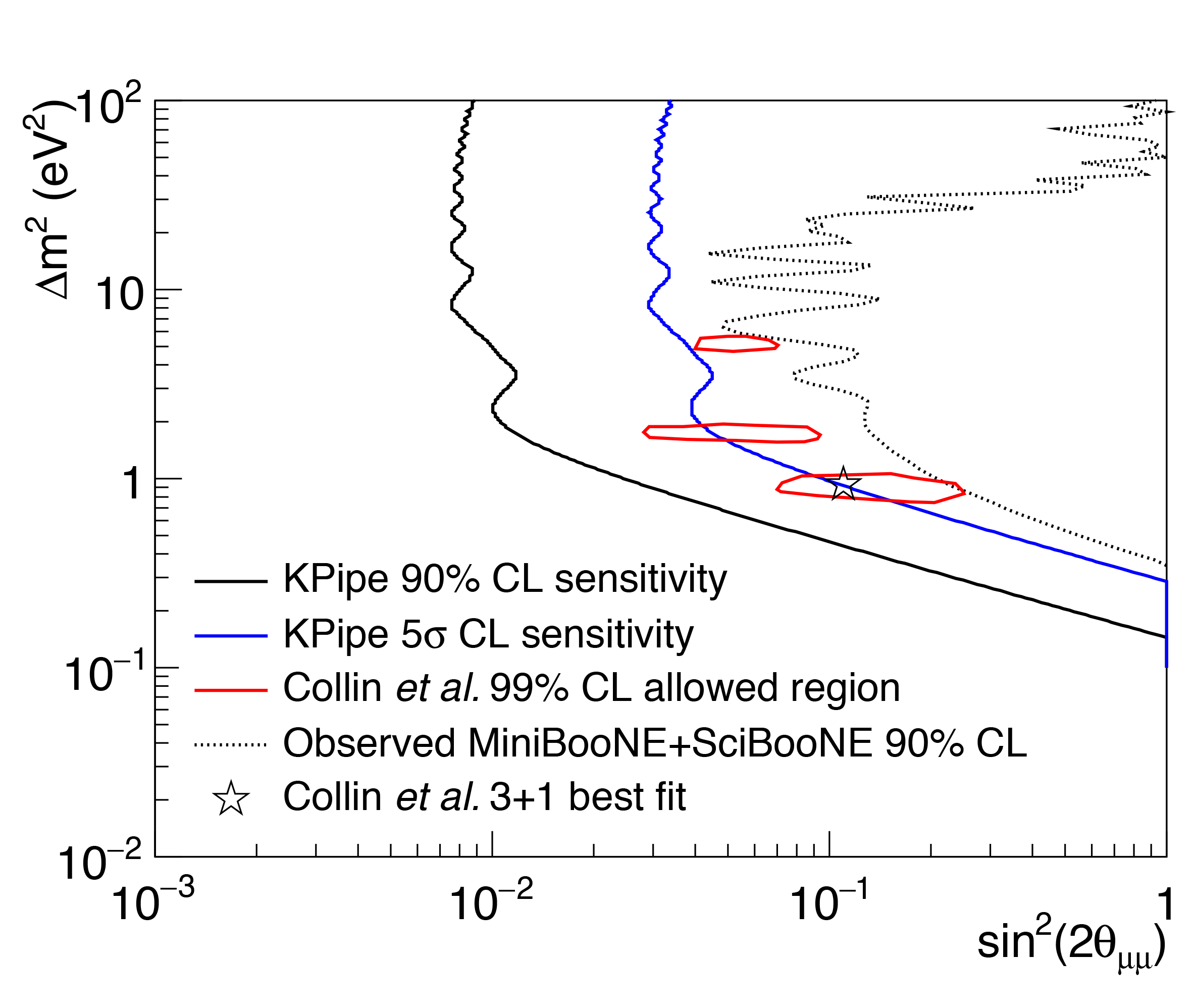}
    \end{minipage}
\caption{Left: The reconstructed oscillation wave as a function of distance along the detector given three different oscillation scenarios. Right: The 90\% and 5$\sigma$ exclusion limits of KPipe after 3 years of running. The current observed excluded region from MiniBooNE and SciBooNE is also shown.}
\label{fig:sens}
\end{figure}

The \numu event rate, given a no-oscillation hypothesis, as a function of distance is calculated numerically using the CC interaction cross section on carbon, kaon production rate, detector live-time, and signal efficiency. Each event is then oscillated according to Equation \ref{disappeq1}, and smeared to incorporate the Gaussian uncertainty on vertex reconstruction and uncertainty on the \numu production location. We present three oscillation scenarios in Figure \ref{fig:sens} (left), which include the statistical uncertainties on the \numu and the CR event rate.  Here, we define $\sin^2(2\theta_{\mu \mu})=4|U_{\mu 4}|^2(1-|U_{\mu 4}|^2)$, to represent the amplitude of the oscillation. It is worth noting that the neutrino energy resolution is inconsequential since the neutrinos are monoenergetic and if we detect a coincident signal that passes the selection criteria, we can be 98.5\% sure that it originated as a KDAR neutrino.

The sensitivity of KPipe is evaluated with a shape-only $\chi^2$ statistic, however, since we do not include any correlated systematic uncertainties, we replace the covariance matrix with the Neyman $\chi^2$ convention (further detail in Ref. \cite{KPipe}). In Figure~\ref{fig:sens} (right), we show the projected 90\% and 5$\sigma$ sensitivity to \numu disappearance given 3 years of running. The red contours represent the 90\% and 99\% current global allowed regions~\cite{collin}. We find that after 3~years of running, KPipe can extend the observed exclusion limits of MiniBooNE and SciBooNE \cite{sci} by an order of magnitude and measure the current global best fit at $\Delta$m$^2 = 0.93$~eV$^2$ and sin$^2$(2$\theta_{\mu\mu}$) = 0.11 to 5$\sigma$.

\section{Conclusion}
The MLF at J-PARC currently houses the world's most intense source of monoenergetic \numu from the decay-at-rest of positive K mesons. KPipe is designed to utilize this source by measuring the \numu charged current interaction rate on carbon nuclei over a distance spanning from 32~m to 152~m in a single detector. In doing so, KPipe can perform a decisive \numu disappearance search in the relevant parameter space for the existence of a sterile neutrino.

After only 3 years of running, KPipe can cover the current global fit allowed \numu disappearance region to 5$\sigma$ and expand the current experimentally excluded region by an order of magnitude. This measurement would help constrain models, which have arisen over the past two decades to explain the short baseline anomalies, and could represent the first decisive sterile neutrino test.

\end{document}